# Tracking the rotation of a birefringent crystal from speckle shift measurement

Kapil K. Gangwar[1,2], Abhijit Roy[2,*], and Maruthi M. Brundavanam[2]

[1]Department of Physics, Indian Institute of Technology Delhi, New Delhi 110016, India

[2]Department of Physics, Indian Institute of Technology Kharagpur, W. B. 721302, India

*abhijitphy302@gmail.com

**Abstract**

The random intensity distribution observed due to the propagation of a coherent beam of light through a scattering medium is known as a speckle pattern. The interaction of the speckles with a birefringent crystal, here, $YVO_4$ results in the e-ray and o-ray speckles. It is observed that the e-ray and o-ray speckles experience an angular shift with the rotation of the crystal and the shift depends on the rotation angle of the crystal. It is also found that the rate of the shift is different for e-ray and o-ray speckles, and is independent of the topological charge of the beam incident on the scattering medium. The observed rate of shift of the speckle pattern is found to be same as that of the e-ray and o-ray beams. It is experimentally demonstrated that the rotation of a birefringent crystal can be tracked very accurately from the speckle shift measurement.

**Introduction**

The interaction of a light beam with a crystal and its applications have been an area of interest of the science community for a long time. The propagation of a beam with sufficient power through a non-linear crystal gives rise to the generation of higher harmonic beams [1]. A high power pulsed laser beam can be used to generate a supercontinuum [2]. The interaction of light with a centro-symmetric crystal can be utilized for the generation and detection of a THz radiation [3]. The distribution of refractive index in a crystal plays a major role in the light-matter interaction. The propagation of a light beam through a birefringent crystal results in the splitting of the beam into the ordinary ray (o-ray) and extra-ordinary ray (e-ray), which is observed due to the double refraction encountered by the light beam during the propagation [4]. The generated e-ray and o-ray are mutually orthogonally polarized: parallel (e-ray) and perpendicular (o-ray) to the principal plane of the crystal, and the double refraction introduces two different refractive indices in these two components. The refractive index of the o-ray has been reported to be independent of the direction of propagation of the beam within the crystal, whereas the refractive index of the e-ray varies according to the direction of propagation [4]. Recently, a lot of effort has been made to understand the interaction between a beam with singularity and a crystal.

An optical vortex (OV) beam is one type of singular beam, which has a spiral wavefront around a singular point, where the phase is undefined, and this type of beam has zero intensity at the center [5]. An OV beam carries orbital angular momentum (OAM), which is defined by the topological charge of the beam. The topological charge is quantified as the number of twists the wavefront of the beam experiences during traversing a distance of one wavelength. The topological charge of a beam is estimated from the z-component of the OAM as, $L_Z = -i\hbar\frac{\partial \phi}{\partial \theta} = -i\hbar l$, where $l$ is the topological charge, and $\hbar$ is the Plank constant divided by $2\pi$ [5]. The unfolding of a uniformly polarized OV beam on propagation through a birefringent crystal has been investigated both theoretically and experimentally from the study of the polarization singularities and Stokes parameters of the output beam [6, 7]. A technique to generate an OV beam by propagating a circularly polarized singular Gaussian beam through a uniaxial crystal has been demonstrated both theoretically and experimentally [8]. It has been observed that the interaction of a vortex free Bessel beam with a chiral crystal can lead to the induction of optical vortices in the Bessel beam [9].

A speckle pattern, which is another type of beam with optical singularity, has been widely used in various applications. The propagation of a coherent beam of light through a random scattering medium leads to the generation of a randomly distributed intensity pattern, commonly known as the speckle pattern [10]. Efforts have also been made to study the light-matter interaction, in case of speckles. The speckles are reported to become temporarily unstable after passing through a liquid crystal cell, which is tuned as a disordered Kerr medium [11]. The interaction between the speckles and a photorefractive crystal connected to an AC power supply has been found to be causing nonstationary polarization modulation on the speckle pattern [12].

The memory effect of a speckle pattern allows it to retain the information of the input beam, and this characteristic along with the two-point intensity correlation based approach have been utilized for different types of imaging and sensing through a random scattering medium [13-15]. Similarly, the intensity cross-correlation, where the intensity correlation between two speckle patterns is estimated to study the change in the speckles due to a perturbation introduced to the system, is another widely used technique for various applications. A technique based on the estimation of speckle displacement, utilizing the intensity cross-correlation based approach, has been reported to be useful for high resolution measurement of the rotation angle of a cylinder [16]. An optical arrangement capable of high-resolution speckle displacement detection can be used for the strain measurement [17]. The velocity of a moving diffused object can be determined exploiting the dynamic properties of the speckles [18]. The phase-shifting interferometry has been extended to the speckles, and several techniques based on the phase-shifting of speckle interferogram have been proposed to measure the vibration amplitude and deformation of a rough surface [19, 20]. This technique can be extended to the displacement and strain measurement, and is also useful for non-destructive testing [21].

In this paper, we have studied the interaction of the speckles with a birefringent crystal, here a $YVO_4$ crystal. The intensity cross-correlation based approach is exploited to study the speckle displacement with the rotation of the crystal. The results are compared to the case, when the scattering medium is not present. The dependence of the rate of shift of the speckles on the topological charge of the input beam is also studied. The experimental details along with the results are presented.

**Theoretical Background**

An OV beam of an integer order, m generated from a 'fork grating' can be written in a cylindrical coordinate system following Ref. [5] as

$$u_m(\rho, \varphi, z) = \sqrt{\frac{\pi}{2}} e^{im\varphi}(-i)^{|m|+1}\exp(\frac{ik}{2z}\rho^2)\frac{Z_R}{z-iZ_R}\sqrt{A}\exp(-A)\left[I_{\frac{|m|-1}{2}}(A) - I_{\frac{|m|+1}{2}}(A)\right] \quad (1)$$

where $I_v$ denoted the modified Bessel function, $\lambda$ is the wavelength of the light, k is the wavenumber of the beam, $Z_R$ is the Rayleigh range, defined as $Z_R = \frac{\pi\omega_0^2}{\lambda}$, and A is defined as $A = \left(\frac{\rho}{z}\right)^2 \frac{kz_R}{4\left(1-i\frac{Z_R}{z}\right)}$.

If the order of the OV beam is fractional, the fractional OV beam of order $\alpha$ can be written as the superposition of infinite number of integer order OV beams following Ref. [5] as

$$u_\alpha(\rho, \varphi, z) = \frac{\exp[i(z+\alpha\pi)]\sin\alpha}{\pi} \sum_{m=-\infty}^{\infty} \frac{u_m(\rho,\varphi,z)}{(\alpha-m)} \quad (2)$$

The characterization of a speckle pattern is performed following the intensity correlation based approach. The shift of the speckles is determined as the shift of the peak of the two-point intensity cross-correlation function of two recorded far-field speckle patterns. The degree of cross-correlation, $\gamma^2(\mathbf{r_1}, \mathbf{r_2})$ of two speckle patterns can be written in terms of the two-point intensity correlation function following Ref. [22] as

$$\gamma^2(\mathbf{r_1}, \mathbf{r_2}) = \frac{\langle \Delta I(\mathbf{r_1}) \Delta I(\mathbf{r_2}) \rangle}{\langle \sigma_I(\mathbf{r_1}) \rangle \langle \sigma_I(\mathbf{r_2}) \rangle} \quad (3)$$

where $\Delta I(\mathbf{r}) = I(\mathbf{r}) - \langle I(\mathbf{r}) \rangle$ is the spatial fluctuation of intensity from its mean value, $\sigma_I$ is the standard deviation of intensity, '$\langle . \rangle$' represents the ensemble average of the variable, and $\mathbf{r_1}$, $\mathbf{r_2}$ are the two spatial position vectors on the transverse plane of two speckle patterns.

**Experimental Details**

The schematic diagram of the experimental setup is shown in Fig. 1. A spatial light modulator, SLM is shined by a linearly polarized beam of light from a He-Ne laser of wavelength of 632.8 nm. In order to introduce different topological charges in the beam, emerging from the SLM, different computer generated holograms (CGHs) are projected on the SLM. The first order of the spatially modulated beam, diffracted from the SLM, which contains the desired topological charge, is then filtered using an aperture, A and the filtered beam is then made to reflect from the mirror, M. The reflected beam is collimated using a combination of lenses $L_1$ and $L_2$ of focal length of 150 mm and 100 mm, respectively. The horizontally polarized collimated beam is made diagonally polarized using a half-wave plate, HWP and is sent through a ground glass plate, GG. The speckles generated from the GG plate are made to pass through a birefringent crystal, here, $YVO_4$ the optic axis of which makes an angle of $45^0$ with the z-axis, which results in splitting of the speckles into e-ray and o-ray speckles. The e-ray and o-ray speckles are filtered using a polarizer, P and are recorded using a charged coupled device or CCD camera. The CCD camera is from Thorlabs having pixel dimension of $1280 \times 1024$ and a pixel pitch of 6.45 μm. In order to study the tracking of the rotation of the crystal, the crystal is rotated at different angles ($\theta$) from $-6^0$ to $+6^0$ in steps of

$1^0$ using a rotational stage, and the e-ray and o-ray speckles are recorded using the camera for each $\theta$.

Initially, the experiment is performed using a beam with zero topological charge i.e. the GG plate is illuminated using a Gaussian beam. To study the dependence of tracking the rotation of crystal on the topological charge of the beam on the GG plate, fork grating patterns are projected on the SLM and beams with topological charge of 0 to 2.0 are generated with an interval of 0.5, and the experiment described in the last paragraph is repeated for each charge.

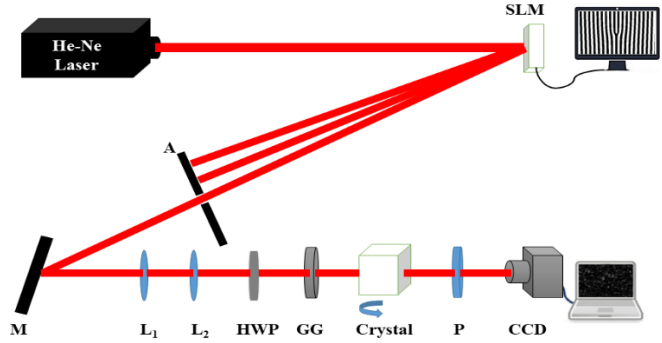

**Fig. 1.** The schematic diagram of the experimental setup.

**Analysis and Results**

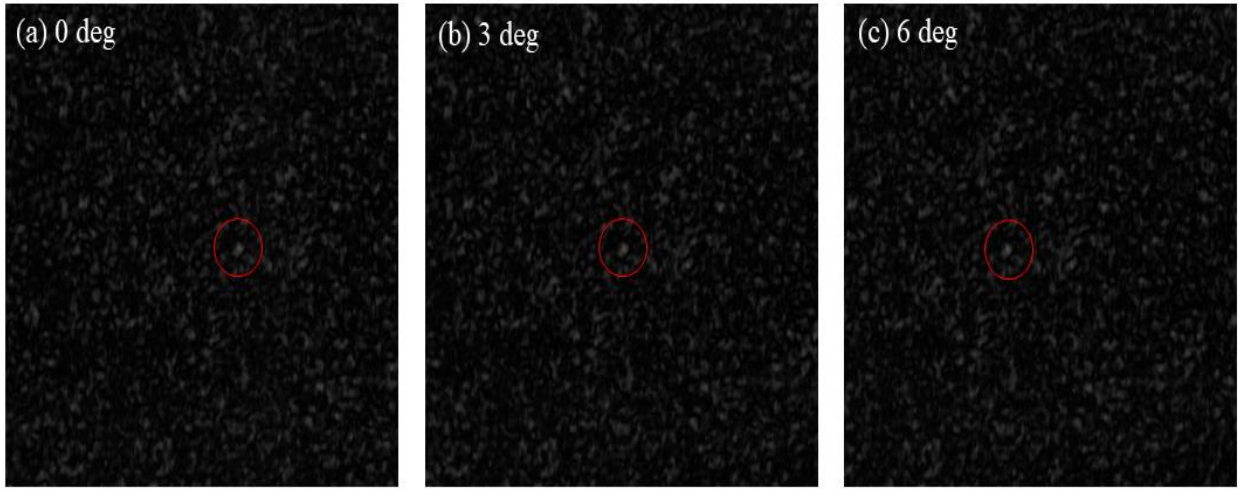

**Fig. 2.** The recorded o-ray speckle pattern for a Gaussian input beam for different $\theta$ of the crystal.

The o-ray speckle patterns recorded for three different $\theta$ of the crystal are shown in Fig. 2, where the shift of the speckles observed with the rotation of the crystal is marked with circles. The shift of the speckles is estimated from the study of the intensity cross-correlation of the speckle patterns recorded at different $\theta$ with the speckle pattern recorded, when $\theta = 0^0$, following Eq. (3), where the ensemble average is replaced with the spatial average under the assumptions of ergodicity and spatial stationarity of the recorded speckle patterns. The amount of shift is determined from the spatial shift of the peak of the degree of cross-correlation. The shift of the speckles is determined for different values of θ and the variation, in case of e-ray and o-ray speckles, for a Gaussian input beam are presented in Figure 3, where a linear shift of the speckles with θ is observed. The slopes of the variation for e-ray and o-ray speckles are found to be 277.35 μm/ degree and 258 μm/ degree, respectively.

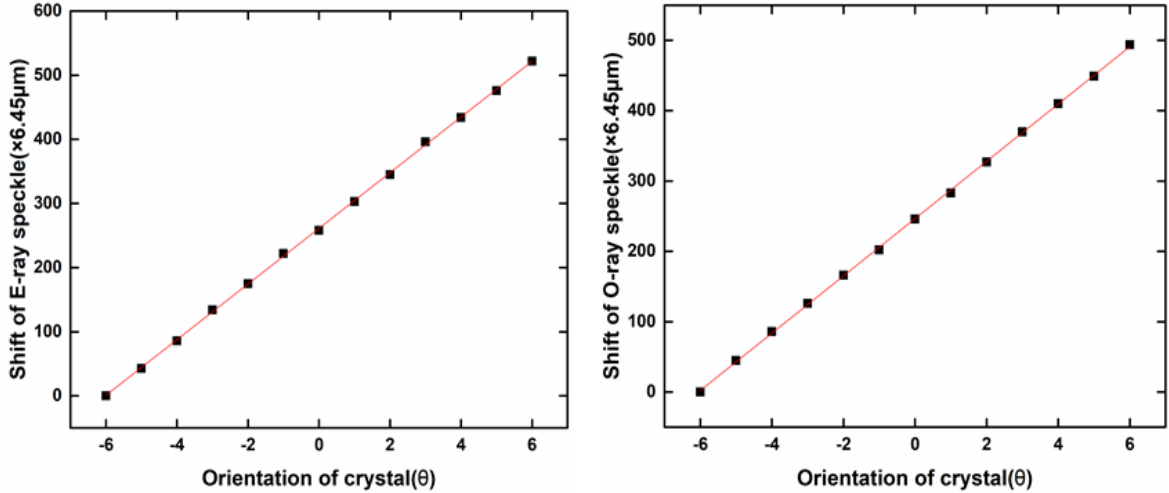

**Fig. 3.** The shift of the e-ray and o-ray speckles with θ for a Gaussian input beam.

In order to study the effect of the presence of a scattering medium on the measurement of the shift, the ground glass is removed and the diagonally polarized Gaussian beam is made to pass through the crystal. The e-ray and o-ray beams for each value of θ are filtered using the polarizer and are recorded using the CCD camera. Although the GG plate is removed, the positions of the crystal, polarizer, and the CCD camera are kept unchanged. The shift of the e-ray and o-ray beams with θ are estimated from the change in the position of center of intensity of the recorded beams for different θ. The shift of the beams with θ in case of a Gaussian input beam in the presence and absence of the scattering medium is shown in Fig. 4, where it can be observed that the presence of the scattering medium does not affect the measurement of the shift.

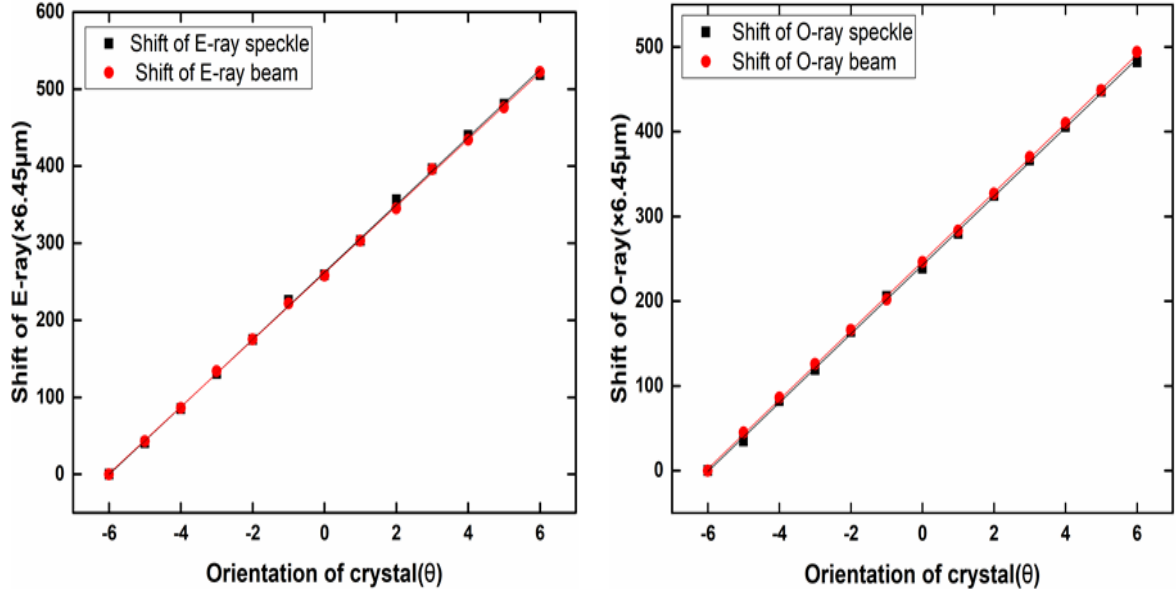

**Fig. 4.** The shift of e-ray and o-ray for a Gaussian input beam in the presence (circular dot) and absence (circular ring) of the GG plate.

The experimental results of the study of dependence of tracking the crystal rotation on the topological charge of the input beam are presented in Fig. 5. The shift of the e-ray and o-ray speckles with θ for different topological charges are presented in Fig. 5(a) and 5(b), respectively, where it can be observed that the variations are linear in nature in both the cases and are also independent of the topological charge of the input beam. The slopes of these variations are found to be same as observed in Fig. 3, and are 43 μm/ degree and 40 μm/ degree in Fig. 5(a) and 5(b), respectively. In this study, the investigation is performed using an input beam with the following integer topological charges: 0.0, 1.0 and 2.0, and the following fractional topological charges: 0.5 and 1.5. The effect of the presence of scattering medium on the measurement of shift for input beams with different topological charges are also studied, and the results for charge 1.0 and 1.5 are presented in Figs. 5(c) to 5(f), where it can be observed that similar to the case of the Gaussian input beam, the presence of the scattering medium does not affect the tracking of the crystal rotation.

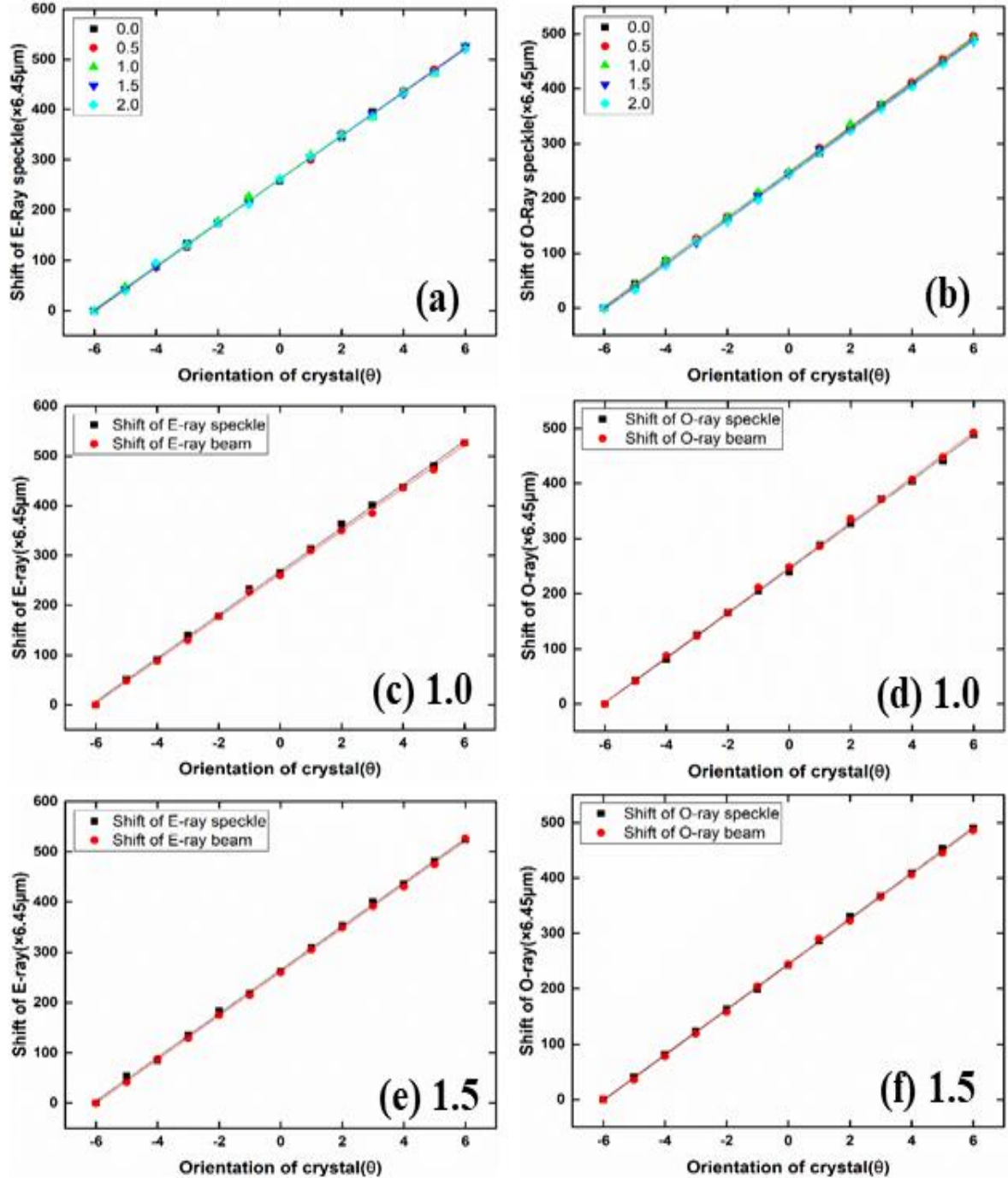

**Fig. 5.** (a - b) The shift of the e-ray and o-ray speckles with θ for different topological charges. The shift of the e-ray and o-ray beam (circular ring) and speckles (circular dot) for the topological charges of the input beam: (c - d) 1.0 and (e - f) 1.5.

The resolution of tracking the crystal rotation through a scattering medium is also determined in case of a Gaussian input beam, as the topological charge of the input beam does not affect the speckle shift measurement. In order to determine the resolution of tracking, the crystal is rotated at an interval of $0.1^0$, and the experimental results of the speckle shift measurement for e-ray and o-ray speckles are shown in Fig. 5, which clearly show that using the proposed technique, the rotation of the crystal equal or more than $0.1^0$ can be easily resolved. Hence, the resolution of tracking the rotation of a crystal through a random scattering medium achieved in the present study is $0.1^0$.

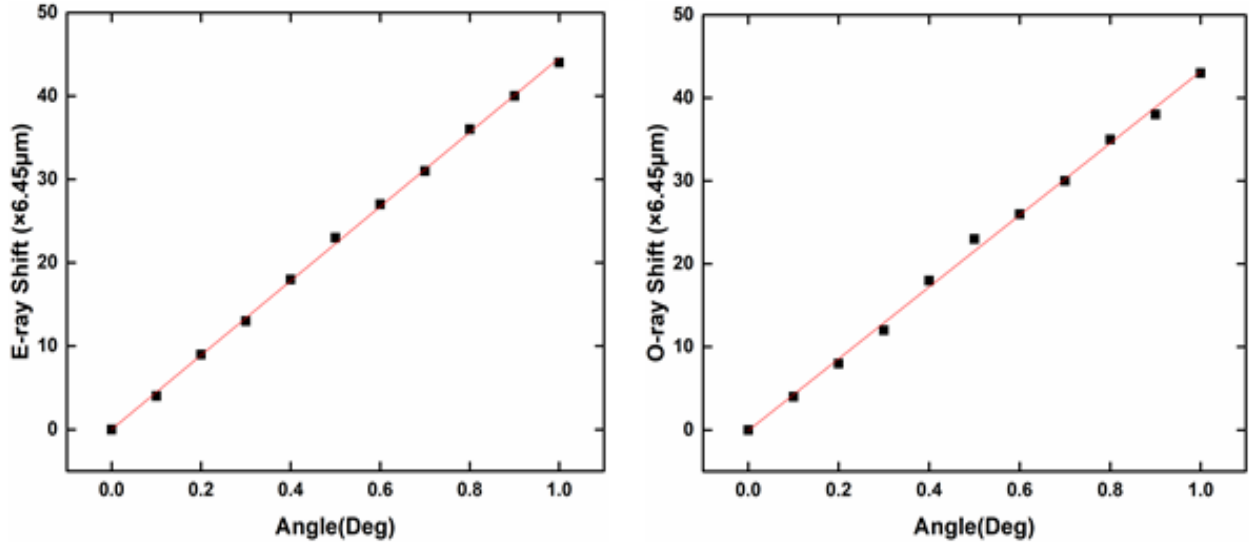

**Fig. 6**. Determination of resolution of tracking the crystal rotation: the shift of the e-ray and o-ray speckle with θ.

## Conclusion

In conclusion, we have demonstrated a technique to track the rotation of a birefringent crystal using a speckle pattern. It is shown that from the speckle shift measurement, it is possible to track the rotation of the crystal, and in our case, the achieved resolution of the tracking is $0.1^0$. It is observed that the rate of the shift in case of e-ray and o-ray speckles are different and is independent of the topological charge of the input beam. It is also found that the presence of the scattering medium does not affect the tracking of the birefringent crystal. The proposed technique can be extended to the case of non-birefringent crystal as well, and as the speckle shift measurement is not affected by the topological charge of the input beam, a Gaussian beam, which is easily available compared to a beam with non-zero topological charge, can be utilized for the tracking. Moreover, as the presence of a scattering medium does not affect the tracking of the crystal rotation, the proposed speckle based approach can be applied to crystals with a wide range of damage threshold due to the fact that the presence of a scattering medium reduces the intensity of the output beam, i.e. the speckles, to a great extent.